\documentclass[manuscript]{aastex}

\usepackage{epsfig}

\slugcomment{}

\shorttitle{}
\shortauthors{Centeno et al.}

\begin{document}


\title{On the magnetic field of off-limb spicules}


\author{Rebecca Centeno}
\affil{High Altitude Observatory (NCAR)\footnote{The National Center for Atmospheric Research is sponsored by the National Science Foundation.}, 3080 Center Green Dr., Boulder CO 80301}
\author{Javier Trujillo Bueno}
\affil{Instituto de Astrof\'\i sica de Canarias, Calle V\'\i a L\'actea S/N, 38205 La Laguna, Spain}
\author{Andr\'es Asensio Ramos}
\affil{Instituto de Astrof\'\i sica de Canarias, Calle V\'\i a L\'actea S/N, 38205 La Laguna, Spain}

\begin{abstract}
 Determining the magnetic field related to solar spicules is vital for developing adequate models of these plasma jets, which are thought to 
play a key role in the thermal, dynamic and magnetic structure of the chromosphere. 
Here we report on the magnetic properties of off-limb spicules in a very quiet region of the solar atmosphere, as 
inferred from new spectropolarimetric observations in the He {\sc i} 10830 \AA\ triplet obtained with the Tenerife Infrared Polarimeter. We have used a novel inversion code for Stokes profiles caused by the joint action of atomic level polarization and the Hanle and Zeeman effects (HAZEL) to interpret the observations. Magnetic fields as strong as $\sim 40$ G were detected in a very localized area of the slit, which could represent a possible lower value of the field strength of organized network spicules.

\end{abstract}


\section{Introduction}

First reported and hand-drawn by the Jesuit Father Angelo \cite{secchi}, solar spicules are often described as chromospheric plasma jets that protude into the solar corona. Since their discovery, they have been observed in a variety of spectral lines, like H$_{\alpha}$, Ca {\sc ii} H and K and the He {\sc i} D$_3$ and 10830 \AA\ multiplets.

\noindent Up until the early 1970s, observational studies of their morphology and dynamics were profuse. The review by \cite{beckers} compiles most of the (observational) spicular knowledge before that time. A 30-year break had to pass before the subject was picked up again (in the early 2000s) and addressed under the scope of high-resolution solar imaging instruments and the new diagnostic capabilities of spectropolarimetry.

Spicules constitute an important consideration for the mass balance of the solar
atmosphere, since they are estimated to carry around a hundred times the mass of the solar wind. 
The average direction of spicules deviates from the vertical, being taller near the poles and 
showing larger local variations in their orientation at lower latitudes. Near the poles, spicules
seem to follow the direction of coronal polar rays, suggesting that their apparent orientation is
closely connected to the surrounding magnetic field topology.

\noindent An essential piece in all spicule model is a magnetic flux tube that reaches from the photosphere all 
the way up into the corona \cite[see][and references therein]{sterling}. An injection of energy into the flux tube is required 
to launch the material and raise it up to heights of several thousand kilometers.
Although models seem to explain some of the observational aspects of solar spicules
(such as approximately constant density and temperature along the lenght of the structure)
they all fail to reproduce some other parameters (such as densities, upward velocities
and lifetimes, etc).

\noindent One of the key impediments is our relatively poor knowledge of the observational properties of
spicules. The increase in spatial resolution offered by some new observing facilities and the further development
and application of the theory of scattering polarization and the Hanle effect for interpreting spectro-polarimetric observations are the keys to recent advances in this field of research \cite[e.g., the review by][]{tb-bangalore}.

Recent Ca {\sc ii} H filtergram observations collected with the SOT broadband imager onboard the solar space telescope Hinode have revealed two distinct types of spicules \citep{depontieu07}. Type I's correspond to our classical image of  spicules: chromospheric plasma structures launched into the corona with lifetimes of $\sim 15$ minutes and oscillation periods between 3 to 7 minutes. Their origin seems to be associated to the leakage of p-modes into the upper atmosphere and subsequent development of shocks that follow the magnetic field lines \cite[see][]{depontieu04}. Type II's, on the other hand, are very dynamic (with lifetimes of less than 2 minutes) thin structures that tend to appear closer to network and plage areas. Speculations point to an  origin related to magnetic field reconnection.

High resolution intensity images like those obtained from Hinode offer an improved view of the morphology and dynamics of these chromospheric structures. However, only with a correct interpretation of detailed spectropolarimetric observations can we aspire to infer their magnetic properties.
This is what drives the present investigation: to establish reliable measured constraints on
some of the physical aspects of spicules, focusing on deciphering the magnetic 
field topology and its behaviour along the length of the spicule. 
In order to pursue such a task we needed high signal-to-noise (S/N) spectropolarimetric measurements
of a suitable chromospheric indicator that is sensitive to the weak magnetic fields that we expect
to be present in these structures.

The first succesful effort in obtaining and interpreting spectropolarimetric measurements to infer the magnetic properties of quiet Sun spicules was reported by \citet{tb-spicules}. These authors measured the Stokes profiles of the He {\sc i} 10830 \AA\ triplet at 2\arcsec.5 off the solar limb and inferred field strengths of 10 G in quiet Sun spicules at an atmospheric height of 2000 km above the visible photosphere. They point out, however, that significantly stronger fields could also be present (as indicated by more sizable Stokes $V$ signals detected during another observing run). The possibility of the existence of magnetic fields
significantly larger than 10 G was also suggested by the He {\sc i} D$_3$ measurements 
of \citet{lopezariste-spicules}, although they observed off-limb spicules in a strongly magnetized plage region. These authors carried out a PCA (Principal Component Analysis) inversion of a series of spectropolarimetric profiles based on a model of optically thin radiative transfer in this multiplet.
More detailed spectropolarimetric observations of spicules in He {\sc i} D$_3$ were carried out by \citet{ramelli-spicules}, who found $B{\approx}10$ G in quiet Sun and $B{\approx}50$ G in more active off-limb areas. 

This paper presents the results obtained from new off-limb spectropolarimetric observations of 
quiet sun spicules in the He {\sc i} 10830 \AA\ multiplet.  Section \ref{observations} describes the observing strategy and the suitability of the selected spectral region. Preliminary results based on the visual inspection of the Stokes profiles (Section \ref{preliminar}) are followed, in Section \ref{inversions}, by a more detailed analysis using the inversion code HAZEL \cite[developed by][]{hazel}, which accounts for the atomic level polarization and the joint action of the Hanle and Zeeman effects. This strategy allowed us to infer the magnetic properties of the observed structures. The presence of Zeeman-induced Stokes $V$ signals in the data gives us a lower limit for the field strength, which is found to be as large as 40 gauss in a localized area of the spectrograph slit.  
\noindent The line-of-sight cancellation of the Zeeman Stokes $V$ signals is illustrated with 
two simple examples of possible scenarios, leading to a discussion on how strong the 
magnetic fields in solar spicules could actually be.

\section{Observations}\label{observations}

Observations were carried out on August 17, 2008 using the Tenerife Infrared Polarimeter 
\cite[TIP;][]{vmp-tip, collados2007} at the German Vacuum Tower Telescope at the Observatorio del Teide (Tenerife,
Spain).
The present version of the TIP instrument provides (almost) simultaneous measurements of the full Stokes vector
for all the spatial points along the spectrograph slit with high spatial ($\sim 0.17$ \arcsec) and spectral samplings ($\sim 11$ m\AA). 

The slit ($0\arcsec.5$ wide and $80 \arcsec$ long) was placed outside the South visible limb and parallel to it, crossing a field of spicules. The 
spectral range spanned from 10826 to 10837 \AA\, covering the He {\sc i} 10830 \AA\ multiplet and a few photospheric and 
telluric lines. With this setup, we carried out a 45-minute long time series in sit-and-stare
mode, with a cadence of 56 seconds. The procedure was repeated for several distances of the slit to the visible limb 
(at 2 and 3 arcseconds). Details about the datasets are listed in Table \ref{tab:datasets}.
Although the seeing was relatively poor, rendering impossible the use of the Adaptive Optics System 
(due to the lack of contrast of the structures inside the limb close to the target), 
the atmospheric conditions were stable enough (0 m/s wind) to ensure that no coelostat 
vibrations were introducing image jitter into the system. The VTT lacks a limb-tracker, so
in order to obtain high quality off-limb observations, very special weather conditions are
required.

The standard data reduction procedure for this instrument was carried out on all the data-sets. 
This includes flatfield and dark current corrections as well as polarimetric calibration \citep[][]{collados99}. 
Each data-set was averaged in time in order to improve the S/N ratio, resulting in the total loss of temporal information. A 33 m\AA\ spectral binning and a 1\arcsec\, spatial average were applied to the data also to increase the S/N.

\begin{table}
\begin{center}
\begin{tabular}{c|cccc}
       & date \& time & length (s) & cadence (s) & height (\arcsec)\\
\hline
1  & Aug 17, 2008 @ 8:04UT & 2692 & 56 & 2 \\
2  & Aug 17, 2008 @ 9:01UT & 2748 & 56 & 3  \\
\end{tabular}
\end{center}
\caption{Details of the two datasets presented in this paper.\label{tab:datasets}}
\end{table}

\section{The polarization of the He {\sc i} 10830 \AA\ triplet}

The He {\sc i} 10830 \AA\ triplet is a powerful diagnostic tool for chromospheric magnetic fields \cite[e.g.,][]{casini2008, tb-bangalore}.
Its particular formation mechanism, triggered by the ionizing effect of the coronal extreme ultra-violet (EUV) illumination,
restricts the absorption and emission processes in these lines to a warped surface at the top of
the chromosphere. The shape of this surface depends not only on the strength and distribution of 
the coronal irradiance but also on the density structure and stratification of the atmosphere \citep[e.g.,][]{avrett, andretta, centenoHe}.

The He {\sc i} 10830 \AA\ transitions share the lower metastable $^3S_1$ energy level in the ortho-helium system.
The transitions between this and the upper $^3P_{2,1,0}$ levels produce two separate spectral signatures, namely
the blue ($^3S_1-^{3}P_0$ at 10829.09 \AA\,) and the red ($^3S_1-^{3}P_1$ and $^3S_1-^3P_2$ blended around 10830.3 \AA\,)
components of the triplet.

The He {\sc i} 10830 \AA\ multiplet is of particular interest for chromospheric diagnostics because its spectral signatures are sensitive to scattering polarization and the Zeeman, Hanle and Paschen-Back effects, which make it a very useful magnetic indicator within a wide range of field strengths \cite[][]{tb-nature, tb-ar-2007}.

\noindent The Zeeman effect is the splitting of the magnetic energy levels of an atom due to the presence of a magnetic field. In the linear regime, this splitting is proportional to the magnetic field 
strength and the Land\'e factor of the level. 
The polarization signals produced by the Zeeman effect are due to the wavelength shifts
between the $\pi$ and $\sigma$ components of the transitions ($\Delta M = 0, \pm 1$). 
When the magnetic field is too weak or absent, the splitting is negligible and no 
polarization signals are generated. Even when the magnetic field is strong enough, if it is unresolved and cancels
out within the spatio-temporal resolution element, the emergent polarization 
signals cancel out too.

Even in the absence of magnetic fields, however, population imbalances among the 
magnetic energy sublevels pertaining to each atomic level can generate polarization signatures \cite[e.g.][]{tb-review}. For instance,
when the number of $\sigma$ transitions per unit volume and time is larger than the 
number of $\pi$ transitions, a linear polarization signal is produced.
Several mechanisms can lead to these population imbalances (atomic level polarization), the key one operating in the Sun being the radiative transitions induced by the anisotropic incident radiation. Due to their height in the atmosphere, chromospheric atoms see the center-to-limb variation of the photospheric continuum radiation. This degree of anisotropy is enough to
produce the so-called optical pumping mechanism, which needs no magnetic field to operate 
and is very effective in generating atomic level polarization if the depolarizing rates from elastic
collisions are low.

\noindent The Hanle effect is just the modification of the atomic level polarization (and of
the resulting emergent polarization signatures) due to the presence of a magnetic 
field that is inclined with respect to the axis of symmetry of the incident radiation field. 
Unlike for the Zeeman effect, the observable effects of atomic level polarization do not tend to cancel out when mixed magnetic polarities are present. Moreover, they are sensitive to very weak magnetic fields.
The range of sensitivity of the upper-level Hanle effect to the field strength is determined by the 
lifetime of the upper energy level of the transition

\begin{equation}
8.79 \times 10^6\, B_{{\rm Hanle}}\, g_l = A_{l}
\end{equation}

\noindent where $B_{{\rm Hanle}}$ is expressed in gauss, $ A_{l}$ (in ${\rm s}^{-1}$) is the Einstein spontaneous emission coefficient and $g_l$ is the Land\'e factor of the level.


\section{Detection of Zeeman-induced Stokes V signals}\label{preliminar}

\begin{figure}
\begin{center}
\includegraphics[angle=0,scale=.40]{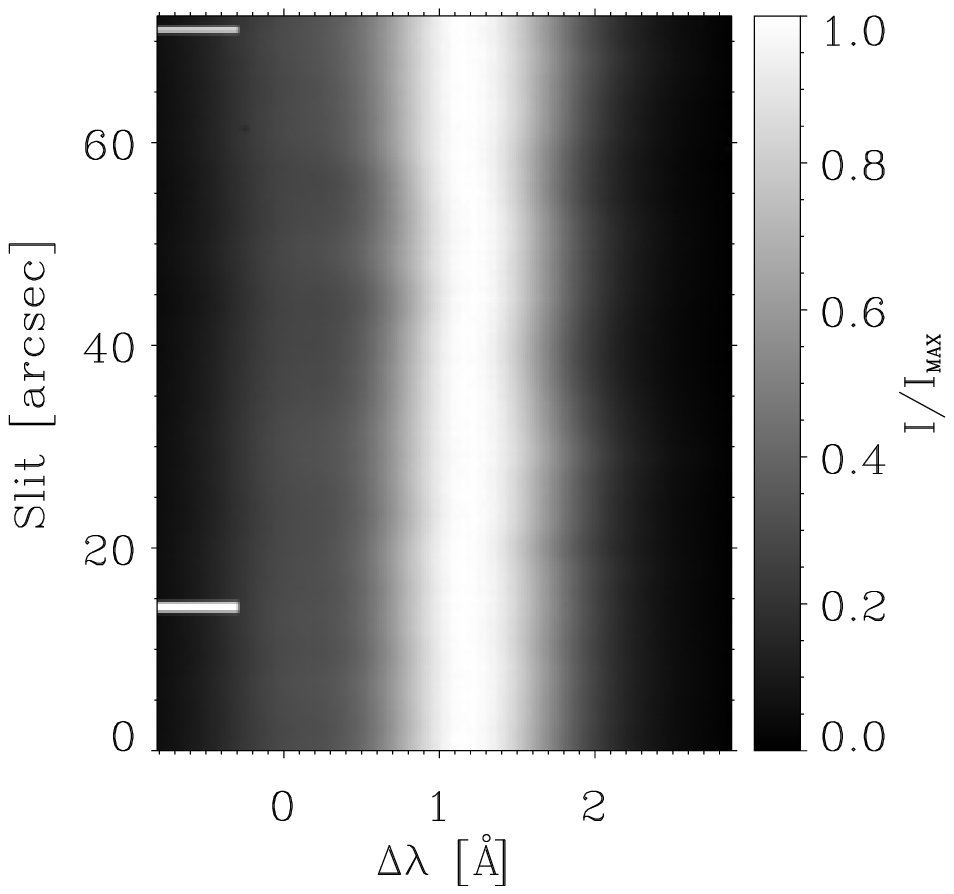}
\includegraphics[angle=0,scale=.40]{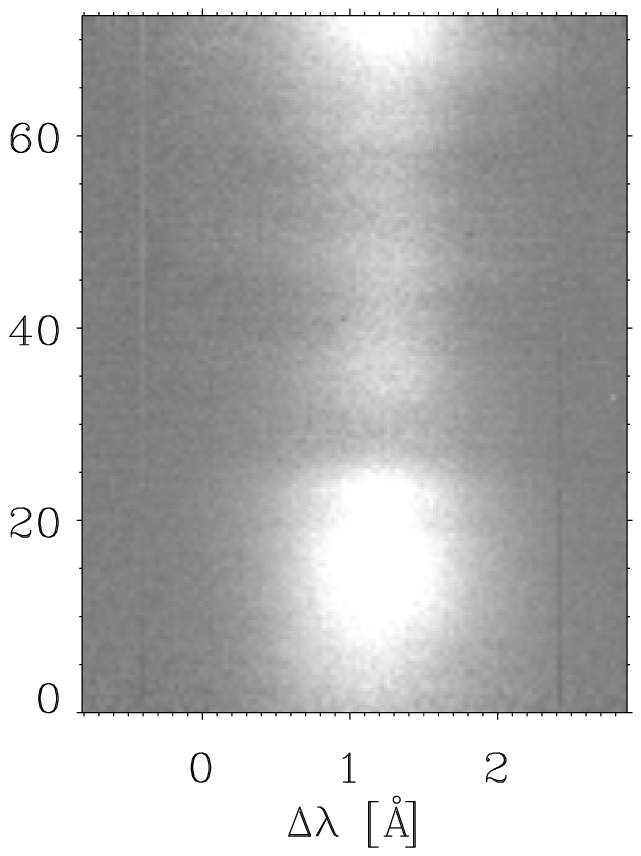}
\includegraphics[angle=0,scale=.40]{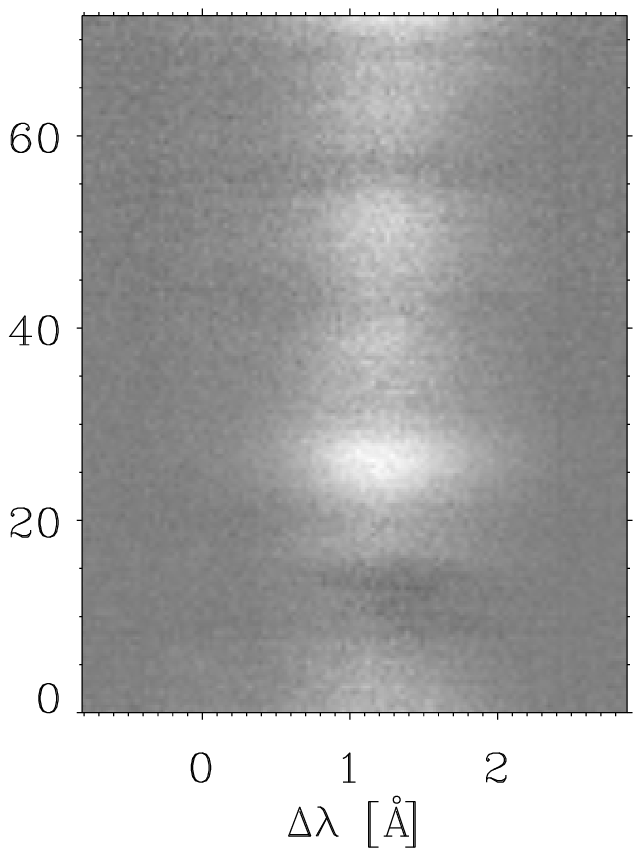}
\includegraphics[angle=0,scale=.40]{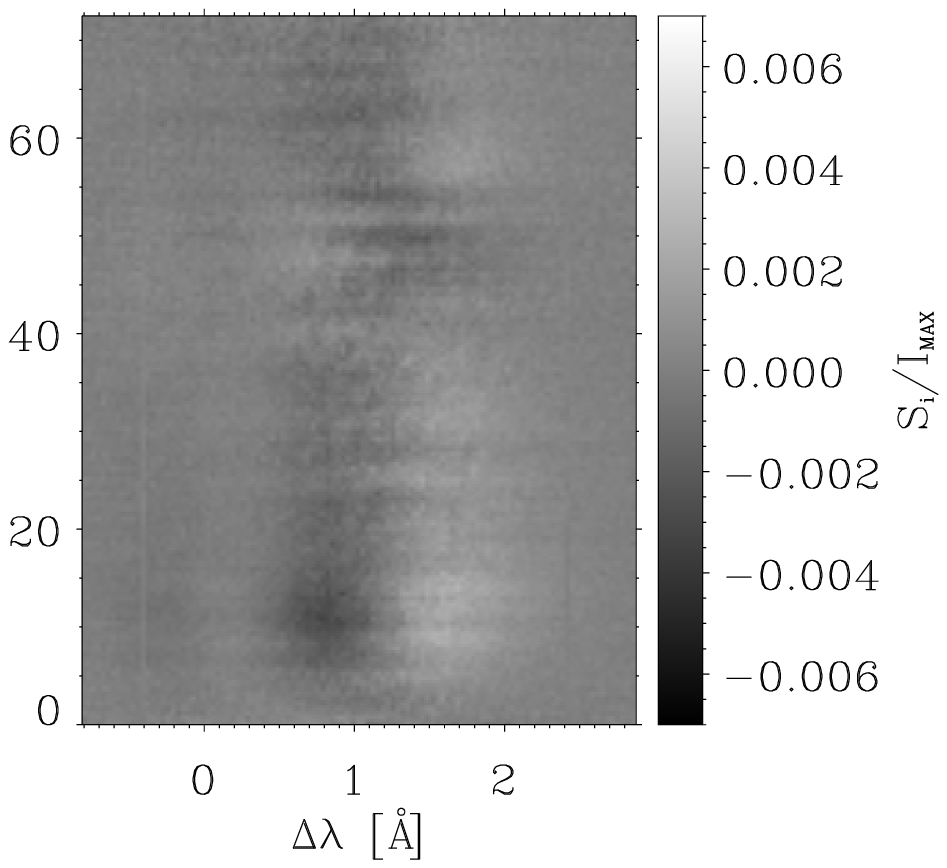}
\caption{Maps of the observed four Stokes parameters (from left to right, $I$, $Q$, $U$, $V$ respectively as a function of wavelength (abscissae) and position along the spectrograph slit. The horizontal line denotes the positions of the profiles extracted for the fits in Fig. \ref{fig:hazelfit}. 
\label{fig:stokes-maps}}
\end{center}
\end{figure}

Fig. \ref{fig:stokes-maps} shows the time-averaged Stokes spectra (from left to right, $I$, $Q$, $U$ and 
$V$, respectively) as a function of wavelength (abscissae) and position along the slit (ordinates), corresponding to data-set \#1.
The brightest vertical strip in the Stokes $I$ panel corresponds to the red component of the He {\sc i} 10830 \AA\ multiplet, while the fainter feature to its left corresponds to the blue component. 
Stokes $I$ provides some physical and thermodynamical information, such as the damping,
the Doppler width (which is related to the combined effects of the thermal and 
non-thermal broadening of the lines), the optical depth, $\tau$ (which gives us information about the number 
of absorbers and emitters along the LOS), and the macroscopic velocity (through the Doppler shift of the line).

\noindent When we combine this information with the linear polarization (Stokes $Q$ and $U$) 
we are able to infer the magnetic field orientation. However, in the Hanle saturation
regime (which is above $\sim 8$ gauss for this multiplet, and remains in such a regime until 100 gauss, approximately), the scattering polarization 
signals are barely responsive to the magnetic field strength, hindering its determination.
Luckyly, the longitudinal Zeeman effect becomes the dominating one shortly above this value, providing nearly continuous diagnosing capabilities throughout the entire domain of field strengths. 
In fact, one of the most striking things that we came across during this particular 
observation was the clear detection of Zeeman-induced Stokes $V$ profiles. As seen in the right panel of Fig. \ref{fig:stokes-maps}, this signal was evident
along the whole field of view of the polarimeter, being stronger in the lower third section of the slit.
This mere fact allowed us to pin down the magnetic field strength.

\noindent A Zeeman-induced Stokes $V$ signal must be due to a net LOS component of the magnetic field, thus we can infer 
the minimum $B_{\rm LOS}$ that explains this feature.
We carried out the exercise of determining the magnetic field component projected onto the line
of sight using only the information from Stokes $V$ and ignoring the linear polarization 
signals. We found that, for the particular case of position 1 (marked by the horizontal line in the lower part of the slit in Fig. \ref{fig:stokes-maps}), the minimum field 
strength that can produce this signal is $\sim 25$ Gauss.
This is just a low-end value of the true magnetic strength, since: (a) this estimation does not take into account the information contained in the linear polarization
profiles, (b) the line-of-sight integration collapses onto one pixel the information coming from regions with differently-oriented magnetic fields, which would result in the partial cancellation of the Zeeman-induced Stokes $V$ signals, and (c) we
are assuming a magnetic filling factor of 1.

\section{Inversions}\label{inversions}

In order to determine the magnetic field strength and other physical magnitudes from
the observations, we carried out the full Stokes inversion of the measured 
spectropolarimetric profiles using the inversion code HAZEL \cite[{\em HAnle and ZEeman Light}, developed by ][]{hazel}.

HAZEL accounts for the physical ingredients and mechanisms operating in the generation
of polarized light in the He {\sc i} 10830 \AA\ triplet, such as optical pumping, atomic level polarization,
level crossings and anticrossings and the fingerprints of the magnetic fields on the spectral signatures through the Hanle and Zeeman effects (with the positions and strengths of the $\pi$ and $\sigma$ 
components calculated within the framework of the Paschen-Back effect theory).

Radiative transfer effects on the emergent intensity and polarization are accounted for by assuming a constant property slab model permeated by a 
deterministic magnetic field $\vec B$. This slab is located at a height $h$ over the visible
solar surface and is illuminated from underneath by the photospheric
continuum. The modification of the radiation anisotropy due to the center-to-limb variation is taken
into account. The optical depth of the slab at the central wavelength of the red blended component, $\tau$, accounts for the integrated number of emitters and 
absorbers along the line of sight, taking care of the collective effect of having 
several spicules along the LOS. 

\noindent HAZEL modifies iteratively the parameters of the slab model in order to produce a set of synthetic Stokes spectra that best match the observed ones in a least-squares sense. This comparison is done in terms of a merit function, $\chi^2$, once the observed Stokes profiles are fed to the code.
First, a global minimization scheme is applied, which explores the whole parameter domain to find an initial model for the Levenberg-Marquardt least squares minimization that follows. After convergence, the inversion yields the model atmosphere that produces the best fit of the synthetic data to the observations.

\begin{figure}
\begin{center}
\includegraphics[angle=0, scale=0.45]{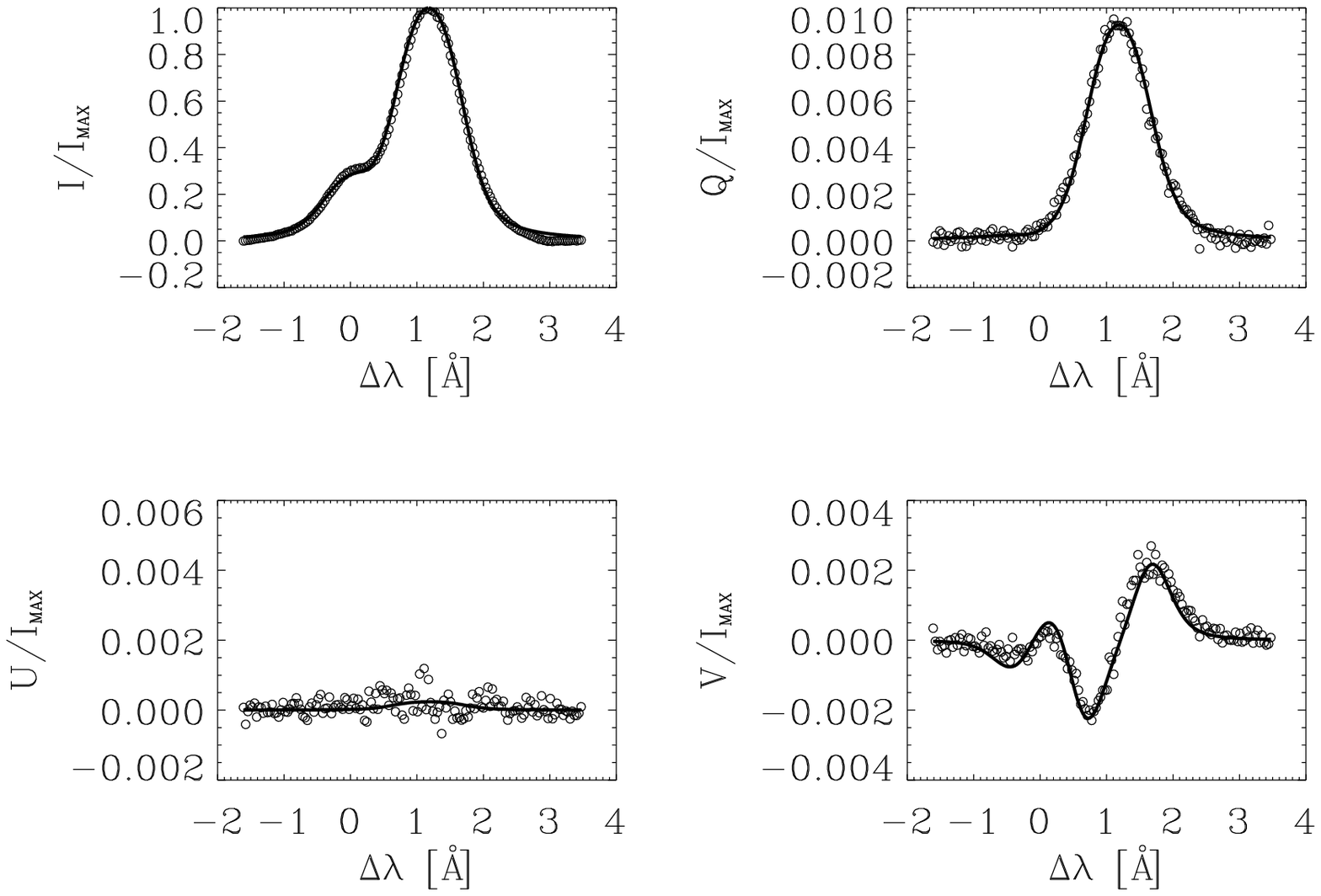}
\includegraphics[angle=0, scale=0.45]{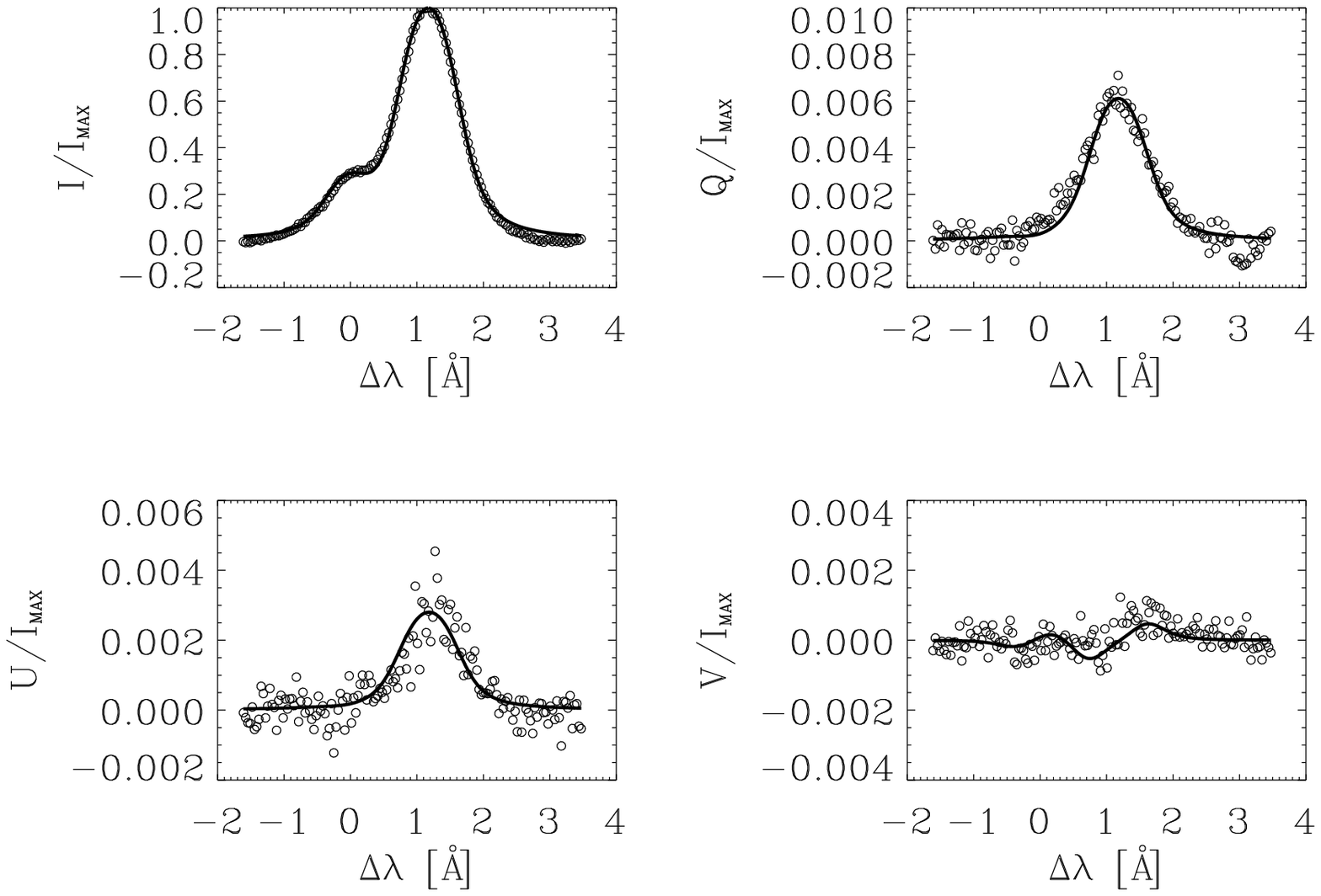}
\caption{Best fit obtained with HAZEL for the Stokes profiles of two separate spatial positions, marked in Fig. \ref{fig:stokes-maps}. The four left-most panels ($B=48$G) correspond to a pixel in the lower part of the slit and the right-most panels ($B=9$G) to a position near the top of the slit. The open circles represent the observed Stokes profiles while the solid line shows the best fit obtained by HAZEL for a certain set of parameters that describe the slab model. These fitting parameters are listed in Table \ref{tab:fits}. \label{fig:hazelfit}}
\end{center}
\end{figure}

With the application of HAZEL we inverted the Stokes profiles for all the positions
along the slit in data-sets \#1 and \#2. Fig \ref{fig:hazelfit} shows two examples of the inversion fits (corresponding to the positions marked in Fig. \ref{fig:stokes-maps}), 
one in which the data show a large Stokes $V$ profile and another one where it is just above the noise level.
The inferred magnetic field strengths are 48 and 9 gauss, respectively. Further results are listed in Table \ref{tab:fits}, where the inclinations are given with respect to the direction of the radiation field anisotropy (i.e. the solar local vertical) and the azimuth has its origin on the LOS, increasing counter clockwise.

\begin{table}
\begin{center}
\begin{tabular}{c|ccccc}
       & B (gauss) & $\theta_B$ ($^{\circ}$) & $\chi_B$ ($^{\circ}$) & $\tau$ & $v_m (km/s)$ \\
\hline
Fit 1  & 48 & 35.4 & 0.14 & 1.43 & -1.14\\
Fit 2  & 9 & 40.2 & -6.7 & 1.49 & -1.21 \\
\end{tabular}
\end{center}
\caption{Inversion results from the fits in Fig. \ref{fig:hazelfit}. Fits 1 and 2 correspond to the left and the right panels of that figure, respectively. \label{tab:fits}}
\end{table}

The magnetic field angles are very well constrained by the observed profiles. Except
for the familiar $180^{\circ}$ ambiguity, a good fit is only possible in the range of a few degrees.
However, the field strength is only well determined when the Stokes $V$ signal is 
present and not noisy. Had there been no Stokes $V$ signal, the model would have been able to produce 
a good fit with any field strength in the range from 5 to 20 G. However, Stokes $V$ was present
along the whole length of the slit, allowing the determination of the field strength.

\begin{figure}
\begin{center}
\includegraphics[angle=0, scale=0.6]{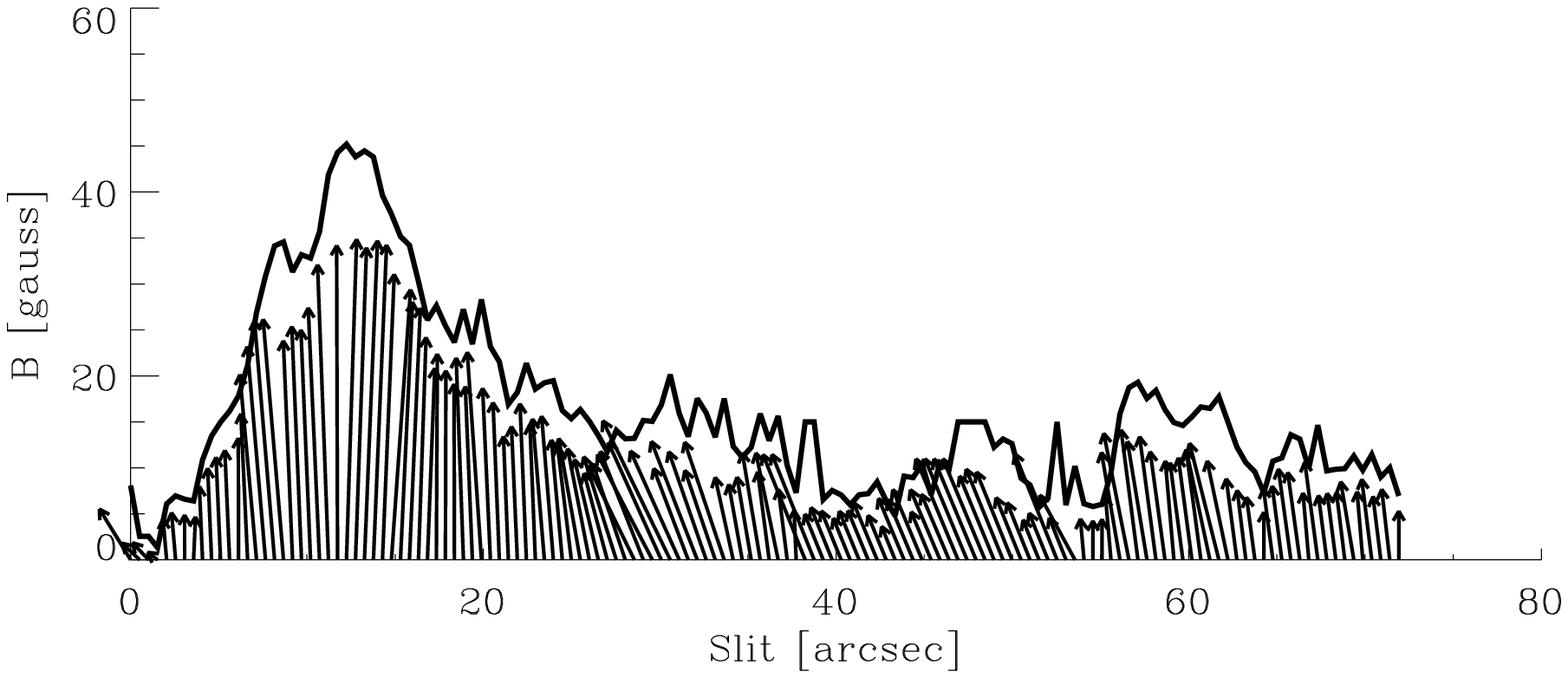}
\includegraphics[angle=0, scale=0.6]{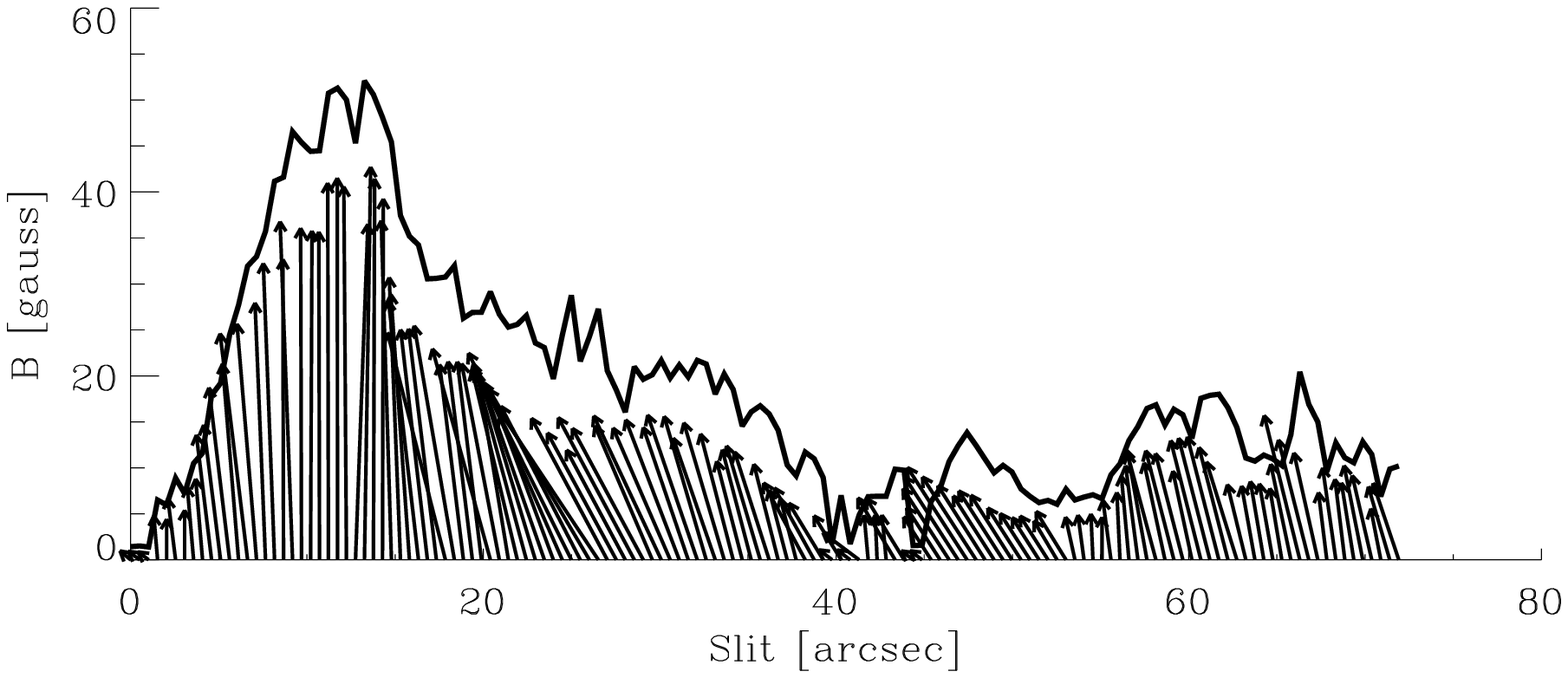}
\caption{Magnetic field inferred by the inversion from the Stokes profiles. The upper panel corresponds to data-set \#1 and the lower panel to data-set \#2. The solid line represents the field strength as a function of the position along the slit, and the arrows show the projection of the magnetic field vector on the plane of the sky \label{fig:sky-projection}}
\end{center}
\end{figure}

The upper panel of Fig. \ref{fig:sky-projection} shows the results corresponding to 
the dataset taken closer to the visible limb (at a distance of 2\arcsec) while the 
lower panel corresponds to the observation in which the slit was placed one arc second 
further away from the visible limb (at 3\arcsec). The solid line shows the inferred field strength. 

\noindent The region with stronger magnetic field corresponds to the section of the slit that shows the largest 
Stokes $V$ signals. The fact that the rest of the areas yield smaller values of the magnetic field 
strength does not necessarily mean that the fields are weaker, but it could be a consequence of the 
cancellation along the LOS of the longitudinal Zeeman effect, since we are integrating along fields of spicules that are combed in different directions. 
On the other hand, in the case of stronger inferred fields, the magnetic field vectors would be 
''teaming up'' to give a net more significant longitudinal component.

The arrows in Fig. \ref{fig:sky-projection} represent the projection on the plane of the sky of the
inferred magnetic field vector. 
If one compares the two panels, it is easy to see that the overall behavior is similar in the two datasets,
implying that they both correspond to positions on the same vertical above the limb.
However, there are some changes in the field strength and in the orientation from one height to another.

\noindent The two datasets correspond to 45-minute long averages and were taken one hour apart from each other.
This means that we have no handle on the temporal evolution of the magnetic fields present in the field of view. For this reason we do not want to extract conclusions on the twist of the magnetic field along the
spicules. We would rather leave this for a future investigation with a different observing strategy.
Spicules show a very rapid evolution, so our results only have a meaning in the context of a relatively long-lived underlying magnetic topology.

\section{Cancellation of the longitudinal Zeeman effect}\label{cancellation}

The mere detection of a Stokes $V$ signal allows us to pin down the magnetic field strength, but 
this will only give us a low-end value for it. 
Let us assume that, along the line of sight, the fields of spicules are combed in different directions. 
Depending on the distribution of the orientations of the magnetic fields, there will be a certain
degree of cancellation of the Zeeman-induced Stokes $V$ signal, that will affect the determination of the longitudinal component of the magnetic field.
Thus, when the measurements do not show any circular polarization signal, 
it could be simply due to cancelling effects along the line of sight. Only in the case that the magnetic field 
lines team up to produce a significant $B_{LOS}$ component, this can result in a non-negligible 
Zeeman-shaped Stokes $V$ signature.

\noindent It is evident from Fig. \ref{fig:stokes-maps} that there are areas along the slit with fairly strong Stokes $V$ signatures (e.g. around $10$\arcsec\, along the slit direction) and regions with very weak circular polarization, suggesting
that, in the latter case, the magnetic field "conspires'' to minimize (or reduce) the net LOS component, resulting in very weak Stokes $V$ profiles.

\begin{figure}
\begin{center}
\includegraphics[angle=0, scale=0.9]{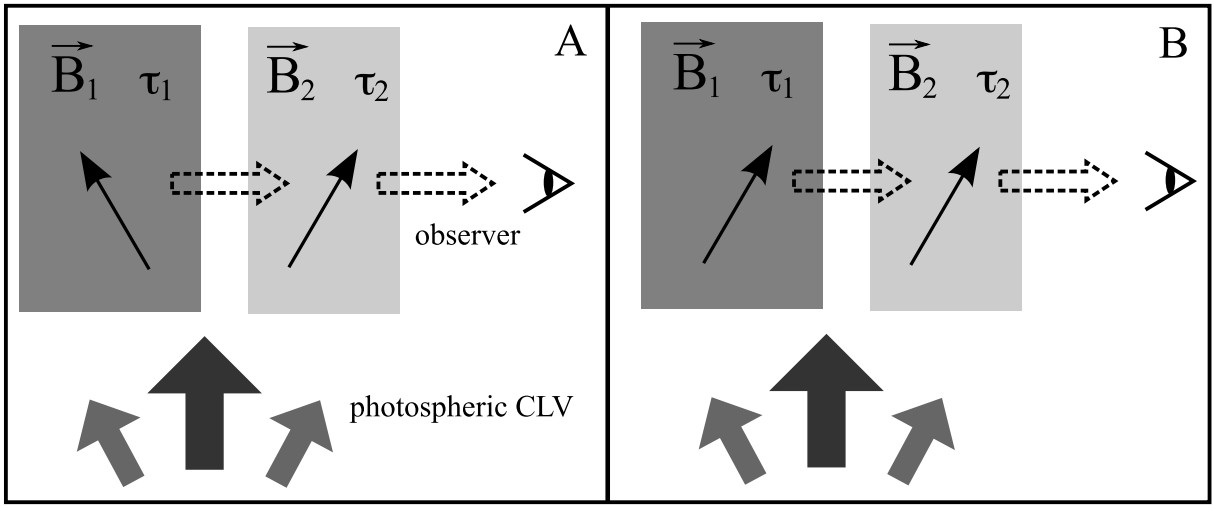}
\caption{Cartoon with two possible scenarios of the magnetic field configuration along the line of sight. In both cases, two slabs contribute to the emergent intensity seen by the observer. In the left panel, the magnetic fields of the two slabs is such that the longitudinal component (along the LOS) cancels out. The right panel represents a case where the field lines in both slabs result in a non-cancelling longitudinal component for the observer.\label{fig:cartoon-slab}}
\end{center}
\end{figure}

We have constructed a simple toy model to illustrate the Zeeman-induced Stokes $V$ cancellation.
Let us consider 2 constant-property slabs of He atoms lying at a height $h$ over the 
solar surface in a 90-degree scattering scenario (see Fig. \ref{fig:cartoon-slab}).
Each slab $j$ (permeated by a deterministic magnetic field $\vec B_j$) is being illuminated 
from below by the photospheric continuum radiation and contributes 
to the emergent intensity at $10830$ \AA\ with an optical depth $\tau_j$ along the LOS.
The total optical depth when the 2 slabs are combined is $\tau = \sum \tau_j = 1.5$ (this corresponds to the average value yielded by the inversions). 
The photospheric illumination incides on both slabs producing atomic level polarization in the He atoms contained in them. When solving the radiative transfer problem, the light emitted by slab \#1 is taken as the background radiation (boundary condition) for slab \#2. Eq. \ref{eq:2slabs1} and \ref{eq:2slabs}, 
taken from \cite[][]{tb-spicules}, show how we have treated the contribution of both slabs to the emergent Stokes profiles:

\begin{eqnarray}
I_{\rm{obs}} =  I_2 + e^{-\tau_2} I_1 ,\label{eq:2slabs1}\\
X_{\rm{obs}} = X_2 + e^{-\tau_2} X_1 - \frac{\eta_{X2}}{\eta_{I2}} \tau_2  e^{-\tau_2} I_1 ,
\label{eq:2slabs}
\end{eqnarray}

\noindent where $I_k$ stands for intensity, $X_k$ represents the Stokes $Q$, $U$ and $V$ generated in the slab $k$,  $\tau_k$ is the integrated optical depth of the corresponding slab, and $\eta$ are the absorption and dichroism components of the propagation matrix. The index $k=2$ refers to the slab that is closest to the observer.

\noindent Two different cases have been considered. In Case A (left panel of Fig. \ref{fig:cartoon-slab}), the magnetic field vectors of the two slabs have a configuration such that the LOS component cancels out almost completely. Case B (right panel of Fig. \ref{fig:cartoon-slab}) assumes that the magnetic fields in both slabs add up producing a non-negligible $B_{LOS}$. 

\begin{figure}
\begin{center}
\includegraphics[angle=0, scale=0.9]{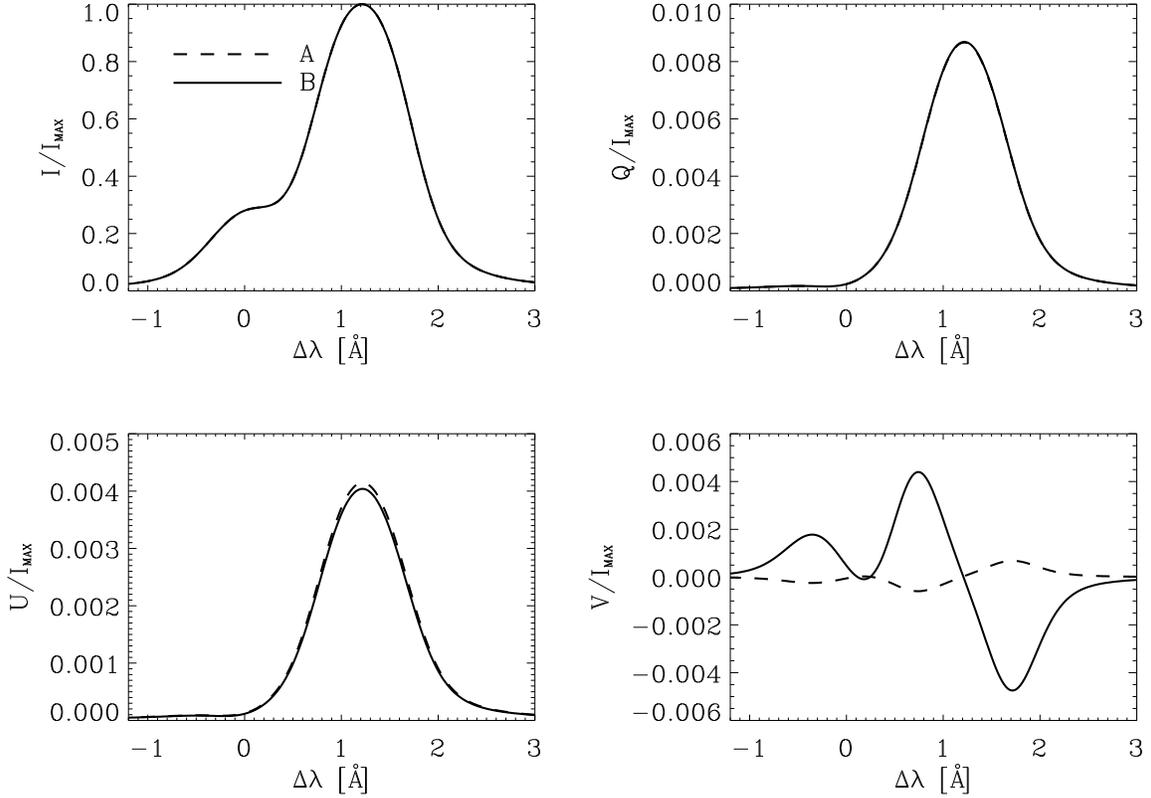}
\caption{Synthetic Stokes profiles generated in the two slab scenario illustrated in Fig. \ref{fig:cartoon-slab}. The dashed line corresponds to Case A, where the magnetic field vectors are such that the longitudinal component cancels out along the LOS. The solid line represents Case B, in which the magnetic field adds up non-destructively to a net circular polarization profile. Note that in both cases, $B=75$ G.\label{fig:synth2slabs}}
\end{center}
\end{figure}

\noindent The sets of parameters used for the synthesis are comprised in Table \ref{tab:slab-models} and the corresponding synthetic Stokes profiles are shown in Fig. \ref{fig:synth2slabs}.
The same optical depth has been given to both slabs, so the contribution to the emergent radiation from slab \#1 is modified due to absorption effects as it travels through slab \#2. For this reason, and taking into account that $\tau$ is not in the optically thin regime, a scenario in which there are 2 identical magnetic field vectors with opposite azimuths, will not cancel out exactly the longitudinal Zeeman signatures.

\begin{table}
\begin{center}
\begin{tabular}{c|cc}
& Case A & Case B\\
\hline
 B$_1$ (gauss)  & 75 & 75\\  
$\theta_1$ ($^{\circ}$) & 45 & 45 \\  
$\chi_1$ ($^{\circ}$)  & -20 & -160\\
$\tau_1$ & 0.75 & 0.75 \\
B$_2$ (gauss)  & 75 & 75\\
$\theta_2$ ($^{\circ}$)   & 32& 32 \\
$\chi_2$ ($^{\circ}$)   & -160 & -160 \\
$\tau_2$ & 0.75 & 0.75\\
\end{tabular}
\end{center}
\caption{Atmospheric parameters for the 2-slab model synthesis for Cases A and B of Figs. \ref{fig:cartoon-slab} and \ref{fig:synth2slabs}.\label{tab:slab-models}}
\end{table}

A variety of scenarios can explain the cancellation of the Stokes $V$ signatures. We have chosen this
2-slab example, with equally distributed optical depths, for the sake of simplicity, thus not implying that
this is the case for this particular observation.
What we intend to transmit to the reader with this simple cartoon
example is that, in the portions of the slit where there is barely any
Stokes $V$ signal, there could still be relatively strong magnetic
fields (of the order of several tens of gauss) whose circular polarization
signatures are cancelling out due to the integration along the line of
sight.

\section{Concluding remarks}

We carried out spectropolarimetric observations of off-limb spicules in the He {\sc i} 10830 \AA\ multiplet at various distances from the visible limb. The measurements show Zeeman-induced Stokes $V$ signals, which are crucial for pinning down the magnetic field strength.

\noindent A number of atmospheric properties were inferred from the observations using the inversion code HAZEL. In particular, the magnetic field vector was determined for all the pixels along the slit and for 
two different heights above the visible limb (on the same vertical), which allowed us to detect twists and gradients in the strength of the field along the length of the spicules. The combination of the He {\sc i} 10830 \AA\ multiplet with the inversion technique has proven
to be very useful to infer the magnetic field of spicules. However, no conclusions can be adventured from these particular datasets, since the measurements at the two different heights correspond to 45-minute averages and were taken around an hour apart from each other. The 
results we infer correspond to averages over this integration time.

Magnetic field strengths as large as 48 gauss were found in a small region of the slit, where the
Zeeman-induced Stokes $V$ signals were particularly large. As argued in Section \ref{cancellation}, 
the inferred field strengths correspond to lower limits of the actual values, since cancellation of the 
longitudinal Zeeman-effect takes place when magnetic field vectors are not aligned.
A possibility is that, in this particular region, we were observing a highly organized patch of network, and the spicules interposed along the LOS contributed constructively to the emergent Stokes $V$ signature.

A statistical analysis of many data-sets of spicules will help to clarify what the typical magnetic field strengths and orientations in these structures are. Simultaneous spectropolarimetric observations of the  He {\sc i} 10830 \AA\ and D$_3$ multiplets would yield more robust results, and, combined with imaging spectroscopy of H$\alpha$ and certain Ca {\sc ii} lines, they would provide a full 3D picture of this aspect of the solar chromosphere.

\section{Acknowledgments}
This work was funded in part by the Spanish Ministry of Science and Innovation through project AYA-2007-63881.

\end{document}